\documentclass[aps,amsfonts,nofootinbib]{revtex4-1}
\usepackage{epsfig}
\usepackage{graphicx}
\usepackage{amsmath}
\usepackage{amsbsy}
\begin{document}
\newcommand{\ee}{\end{equation}}
\newcommand{\bb}{\begin{equation}}
\newcommand{\eqb}{\begin{eqnarray}}
\newcommand{\eqf}{\end{eqnarray}}
\def\sigmavec{\mbox{\boldmath$\sigma$}}
\def\x{\mathbf{x}}
\def\p{\mathbf{p}}
\def\ho{{\mbox{\tiny{HO}}}}
\def\sc{\scriptscriptstyle}
\newcommand{\1}{{\'{\i}}}
\def\sigmavec{\mbox{\boldmath$\sigma$}}
\def\nablavec{\mbox{\boldmath$\nabla$}}
\def\thetavec{\mbox{\boldmath$\theta$}}
\def\nc{{\mbox{\tiny{NC}}}}
\def\openone{\leavevmode\hbox{\small1\kern-3.6pt\normalsize1}}
\title{Spin Non-commutativity and the Three-Dimensional Harmonic Oscillator }

\author{H. Falomir}
\email{falomir@fisica.unlp.edu.ar}
\affiliation{IFLP - CONICET and Departamento de  F\'{\i}sica, Facultad de Ciencias Exactas de la UNLP, \\
C.C.\ 67, (1900) La Plata, Argentina}
\author{J. Gamboa }
\email{jgamboa55@gmail.com}
\affiliation{Departamento de  F\'{\i}sica, Universidad de  Santiago de
  Chile, Casilla 307, Santiago, Chile \\ and \\
Facultad de  F\'{\i}sica, Pontificia Universidad  Cat\'olica de Chile, Santiago, Chile}
\author{M. Loewe}
\email{mloewe@fis.puc.cl}
\affiliation{Facultad    de   F\'{\i}sica,    Pontificia   Universidad
  Cat\'olica de Chile, Santiago, Chile}
\author{F. M\'endez}
\email{fernando.mendez.f@gmail.com}
\affiliation{Departamento de  F\'{\i}sica, Universidad de  Santiago de
  Chile, Casilla 307, Santiago, Chile}
\author{J. C. Rojas}
\email{jurojas@ucn.cl}
\affiliation{Departamento  de F\'{\i}sica, Universidad  Cat\'olica del
  Norte, Antofagasta, Chile}
\begin{abstract}
  A three-dimensional harmonic  oscillator with spin non-commutativity
  in  the  phase  space  is  considered.  The  system  has  a  regular
  symplectic structure  and by using  supersymmetric quantum mechanics
  techniques,  the ground state  is calculated  exactly. We  find that
  this state is infinitely  degenerate and it has explicit spontaneous
  broken symmetry.  Analyzing the  Heisenberg equations, we  show that
  the total angular momentum is conserved.
\end{abstract}
\maketitle
In  the last ten  years the  implications of  non-commutative geometry
\cite{nikita} has been intensively investigated as a way, for example,
to develop computational techniques for understanding problems that go
beyond perturbation theory. A  particularly interesting example is the
quantum Hall effect  where -- due to the fact  that the magnetic field
is strong  -- non-perturbative techniques are  required.  Besides, the
study  of  motion  of  charged  particles in  strong  magnetic  fields
\cite{landau,wil} is  an interesting mathematical  problem because, as
it  is  known, the  commutator  of  the  momenta (more  precisely  the
covariant derivatives),  is different from zero. In  other words, this
is a genuine example of a non-commutative geometry problem.

Some  years ago,  Nair  and Polychronakos  \cite{np} studied the noncommutative
harmonic oscillator, {\it i.e.}  a system described by the Hamiltonian
\bb
{\hat H} = \frac{1}{2} \left( {\bf p}^2 + {\bf r}^2\right), \label{1}
\ee
and the deformed commutators \cite{some} given by
\eqb
\left[{\hat x}_i,{\hat x}_j\right] &=& i \theta_{ij}, \nonumber
\\
\left[{\hat p}_i,{\hat p}_j\right] &=& i B_{ij}, \label{2}
\\
\left[x_i,p_j\right] &=& i\delta_{ij}.  \nonumber
\eqf
with  $\{i,j\}\in\{1,2,3\}$. This problem is exactly soluble  and presents a fixed
point.

In  three   dimensions,  for  example,  if   we  choose  $\theta_{ij}=
\epsilon_{ijk}\theta_k$   and  $B_{ij}   =   \epsilon_{ijk}B_k$,  with
$\theta_i,B_j$  constant vectors,  then  the  determinant of  the  symplectic
matrix $\Omega_{ab}=[\xi_a,\xi_b]$  -- with $\{ \xi_a \} \equiv({\bf x},{\bf p})$
--  turn  out  to be  $\mbox{det}(\Omega)=-(1-{\thetavec}\cdot{\bf
B})^2$ which, therefore,  vanishes for ${\thetavec}\cdot {\bf  B} = 1$
\cite{np}.   Alternatively,  it is  possible  to  perform a  coordinate
transformation  in the  phase space  in order  to get  a  well behaved
symplectic matrix,  but in such a case,  the coordinate transformation
turn  out to be  singular in  the parameter  space $\{\theta_i,B_k\}$,
again for ${\thetavec} \cdot {\bf B} = 1$.

An interesting  problem, 
occurs when
the spatial variables are mixed with the spin variables in such a way that
the antisymmetric tensors in the right hand sides of Eqs.\ (\ref{2}) are given by
\bb
\theta_{ij}       =\theta^2      \epsilon_{ijk}       {\hat      s}_k,
\,\,\,\,\,\,\,\,\,\,\,\,\,\,\,  B_{ij} =\kappa^2  \epsilon_{ijk} {\hat
  s}_k, \label{3}
\ee
where  $\theta$   and  $\kappa$   are  length  and   momentum  scales,
respectively  (in natural  units with  $\hbar=1$), and  $\hat{s}_k$ are
spin matrices.

The  deformation  of  the  second commutator  in  (\ref{2})  with
$B_{i j}$ specified in  (\ref{3}) is a kind of non-relativistic \cite{gosh}
version  of  the Snyder-Yang  algebra  \cite{snyder,yang}.  Indeed,  the
choice   (\ref{3})   corresponds  to the reduction of  $M_{\mu   \nu}  =   \frac{i}{2}
[\gamma_\mu,\gamma_\nu]$
to just $\sim i \epsilon_{ijk}\sigma_k$.

The complete set of consistent commutation relations
\begin{equation}
\label{4}
\begin{array}{ll}
\left[{\hat  x}_i,{\hat x}_j\right]  = i\theta^2  \epsilon_{ijk} {\hat
  s}_k   &   \left[{\hat   p}_i,{\hat   p}_j\right]   =   i   \kappa^2
\epsilon_{ijk} {\hat s}_k
\\
\left[{\hat  x}_i,{\hat p}_j\right]  = i(\delta_{ij}  +  \kappa \theta
\epsilon_{ijk} {\hat s}_k) &
\left[{\hat s}_i,{\hat s}_j \right] = i\epsilon_{ijk}{\hat s}_k
\\
\left[ {\hat  x}_i, {\hat  s}_j \right] =  i \theta\epsilon_{ijk}{\hat
  s}_k  &   \left[  {\hat  p}_i,   {\hat  s}_j  \right]  =   i  \kappa
\epsilon_{ijk}{\hat s}_k
\end{array}
\end{equation}
can be  explicitly realized in terms of canonical
variables through the shift
\eqb
{\hat x}_i \to {\hat x}_i &=& x_i + \theta s_i, \nonumber
\\
{\hat p}_i \to {\hat p}_i &=& p_i + \kappa s_i, \label{5}
\\
{\hat s}_i \to {\hat s}_i &=& s_i = \frac{1}{2} \sigma_i, \nonumber
\eqf
where  $(x_i,p_j)$ obey  the  standard
Heisenberg  algebra and  the  identification $s_i=\frac{1}{2}\sigma_i$
corresponds  to spin-$1/2$, the  situation we  shall consider  in this
paper.

These noncommutative variables give rise to the symplectic matrix
\bb
\begin{pmatrix}
[\hat{x}_i,\hat{x}_j] & [\hat{x}_i,\hat{p}_j] & [\hat{x}_i,\hat{s}_j]
\\
[\hat{p}_i,\hat{x}_j] & [\hat{p}_i,\hat{p}_j] & [\hat{p}_i,\hat{s}_j
\\
[\hat{s}_i,\hat{x}_j] & [\hat{s}_i,\hat{p}_j] &[\hat{s}_i,\hat{s}_j]
\end{pmatrix}
\ee
which is regular, being its determinant a nonvanishing constant
(independent of $\theta$ and $\kappa$)\footnote{The same is also true for,
at least, the model with spin 1.}. This  is  a  very  interesting  peculiarity  of  the  oscillator  with
non-commutativity  of  spin, which  justifies  a more  careful
analysis of their properties.

Thus, following  the prescription in Eq.\ (\ref{5}), the  Schr\"odinger equation is
\bb {\hat H}  (x_i +\theta \frac{\sigma_i}{2}, p_i  + \kappa
\frac{\sigma_i}{2}) |\psi  (t)> =  i \partial_t |\psi  (t)>. \label{6}
\ee

The model based on the harmonic oscillator we are considering   in the present
paper is then defined by the Hamiltonian \cite{parmi1}
\eqb
{\hat H}  &=& \frac{1}{2} \left( {\bf  {\hat p}}^2 +  {\hat {\bf r}}^2
\right)\nonumber
\\
&=&  \frac{1}{2} \left(  {\hat p}_i  +i{\hat x}_i  \right)\left( {\hat
    p}_i -i{\hat x}_i \right) + \frac{3}{2} \nonumber
\\
&=& A^{\dagger}_i A_i + E_0, \label{7}
\eqf
where $E_0$ is the energy of the ground state and
\eqb
A_i &=&  \frac{1}{\sqrt{2}}\left({\hat p}_i -i{\hat x}_i \right), \nonumber
\\
A^{\dagger}_i  &=&  \frac{1}{\sqrt{2}}\left({\hat  p}_i  +i{\hat  x}_i
\right) \label{8}
\eqf
are  the  analogous  of  the  creation  and destruction  operators  of  the  usual
commutative case, but now they satisfy the algebra
\eqb
\left[   A_i,   A_j\right]   &=&   \frac{i}{2}   (\kappa   -i\theta)^2
\epsilon_{ijk} s_k, \nonumber
\\
\left[   A^\dagger_i,  A^\dagger_j\right]   &=&   \frac{i}{2}  (\kappa
+i\theta)^2 \epsilon_{ijk} s_k, \label{9}
\\
\left[   A_i,  A^\dagger_j\right]  &=&   \delta_{ij}  +   i  (\kappa^2
+\theta^2) \epsilon_{ijk} s_k, \nonumber
\eqf
which   is  smooth  in   the  commutative   limit,  {\it   i.e.}  when
$\theta,\kappa \to 0$.

To further discuss the  physical content of this generalization of the harmonic oscillator with such a nonconventional algebraic  structure,  it  is  useful  to  consider  the  supersymmetric  version
associated to  (\ref{7}) and study, for  example, the ground state of the model.

In principle this is a  direct calculation: firstly we redefine the zero energy level by subtracting the constant $E_0$ to the Hamiltonian, {\it i.e.} we change ${\hat H}\rightarrow{\hat H} + E_0$ so that (9) changes into
\bb
\hat{H}=  A^{\dagger}_i A_i, \label{10}
\ee
which is a positive semi-definite  operator.

Following    the    supersymmetrization    procedure    proposed    in
\cite{gz,gozzi}  (see also \cite{nos}),  the supercharges  are defined
as follows
\eqb
A_i\,  \rightarrow \,\,\,  Q&=&\, A_i  \psi_i =  A_i  \sigma_i \otimes
\sigma_- \equiv A\otimes \sigma_-, \nonumber
\\
A^\dagger_i \to Q^\dagger &=& A^\dagger_i \psi^\dagger_i = A^\dagger_i
\sigma_i \otimes \sigma_+ \equiv A\otimes \sigma_+, \label{11}
\eqf
where  $\sigma_\pm  =\frac{1}{2}  (\sigma_1  \pm  i  \sigma_2)$
fulfil $\sigma_\pm^2=0$ and, therefore, $\psi_i^2 =0={\psi_i^\dagger}^2$.

The supersymmetric Hamiltonian is then defined as
\eqb
H &=&  \frac{1}{2} \{Q^\dagger, Q\} = \frac{1}{2}  A^\dagger A \otimes
\frac{\openone_{2\times 2 }+\sigma_3}{2} + \frac{1}{2} AA^\dagger \otimes
\frac{\openone_{2\times 2 }-\sigma_3}{2},
\label{12}
\\
&=&  \frac{1}{2}\left( {\bf  p}^2 +  {\bf r}^2  + 3  \theta ~\sigmavec
  \cdot  {\bf r} +  3 \kappa  ~\sigmavec \cdot  {\bf p}  + \frac{9}{4}
  (\theta^2 +  \kappa^2) \right)\otimes \openone_{2\times  2 } -\left(
  \frac{3}{2} + \sigmavec \cdot {\bf L}\right) \otimes \sigma_3,
 \label{122}
\eqf
which commutes with the supercharges
\begin{equation}
\left[H, Q \right] =0= \left[ Q^\dagger, H \right]. \nonumber
\label{13}
\end{equation}
This is the standard supersymmetric algebra.

The last term of (\ref{122}) is the spin-orbit coupling that emerges from the
usual  three-dimensional supersymmetry,  while $\sigmavec\cdot{\bf  r}  $ and
$\sigmavec\cdot{\bf  p}$ correspond  to magnetic  dipolar  and Dresselhaus
\cite{dressel} interactions respectively.

In order to  find the ground states one note  that the supercharge $Q$
annihilates the vacuum, namely
\bb
Q\, \Psi_0 =0, \label{15}
\ee
where $\Psi_0$  denotes the ground state. The previous equation
becomes
\bb
A_i \sigma_i \otimes \sigma_-\, \Psi_0=0
\ee
or, more explicitly,
\bb
\begin{pmatrix}
0& 0
\\
A_i \sigma_i& 0
\\
\end{pmatrix}
\begin{pmatrix}
\Psi^I_0
\\
\Psi^{II}_0 \end{pmatrix} =0 \Rightarrow A_i \sigma_i\, \Psi^I_0 =0,
 \label{di}
\ee
with $\{\Psi^I_0,\Psi^{II}_0\}$ two-components spinors.

In matrix form, Eq. (\ref{di}) writes as
\bb
\left[ \sigmavec \cdot ({\bf p} - i {\bf x}) + M\right] \Psi^I_0 =0, \label{16}
\ee
where  $M= \frac{3}{2} (\kappa - i \theta)$ is a complex number.

Equation (\ref{16}) is solved by functions of the form
\bb
\Psi^I_0 =  e^{-\frac{{\bf x}^2}{2} +i {\bf k}\cdot{\bf  x}} \,u ({\bf
  k}), \label{17}
\ee
where $u ({\bf k})$ is a constant two-components spinor satisfying the
conditions
\bb
\left( \sigmavec. {\bf k} + M \right) u({\bf k})=0, \label{177}
\ee
with  ${\bf k} = {\bf k}_R + i {\bf k}_I\in \mathbb{C}^3$, a complex vector.

The  condition (\ref{177}) implies that ${\bf k}^2=M^2$ and then
\eqb
{\bf k}^2_R - {\bf k}^2_I &=& \frac{9}{4} (\kappa^2 -\theta^2), \nonumber
\\
{\bf k}_R . {\bf k}_I &=&- \frac{9}{4} \, \kappa\, \theta.
\eqf
These equations are the same we found in a previous work \cite{nos}.

Notice that the conditions on $\mathbf{k}=\mathbf{k}_R+i \mathbf{k}_I$ determine only the square ${\bf k}^2$. Then, there is a continuous of degenerate minima which can be obtained from one of them by a unitary transformation given by  $3\times 3$ complex  matrices satisfying $U^t U = \mathbf{1}_3$.   Near the identity, $U=e^{iM}\simeq \mathbf{1}_3+ i M$, the previous condition requires $M^t=-M$. Therefore, the Lie algebra of this group is the space of complex and antisymmetric $3\times 3$ matrices. This corresponds to  the Lie algebra of the complexification $\overline{SO(3)}$ of $SO(3)$, which has the covering group $SL(2,\mathbb{C})$. This Lie algebra is generated by $3\times 3$ matrices $\left\{X_i, \bar{X}_i, i=1,2,3 \right\}$ which satisfy the commutation relations $[X_i,X_j]=i \epsilon_{ijk}X_k$, $[X_i,\bar{X}_j]=i \epsilon_{ijk}\bar{X}_k$ and $[\bar{X}_i,\bar{X}_j]=- i \epsilon_{ijk}X_k$.

Thus, for a given $\mathbf{k} \in \mathbb{C}^3$,  the solution for the ground state in Eq.\ (\ref{17}) has the symmetry corresponding to rotations around this given direction and, therefore, its little group is $SO(2)\times SO(2)$, generated by $\mathbf{k}\cdot \mathbf{X}$ and $\mathbf{k}\cdot \mathbf{\bar{X}}$. So, the symmetry of the Hamiltonian $\overline{SO(3)}$ is broken down to $SO(2)\times SO(2)$, and the transformations which move from one ground state to another are elements of the quotient group $\overline{SO(3)}/ \left(SO(2)\times SO(2)\right)$.

Let  us  discuss  now   the problem  concerning  to  the  dynamical
evolution of the  operators ${\hat x}, {\hat p}$  and ${\hat s}$. This
evolution  is   determined  by  the  Heisenberg   equations  for  such
operators, {\it i.e.}
 \begin{equation}
{\dot {\hat x}}_i = \frac{1}{i} [ {\hat x}_i, H],~~~~~~
{\dot {\hat p}}_i = \frac{1}{i} [ {\hat p}_i, H],~~~~~~
{\dot s}_i = \frac{1}{i} [ s_i, H],
 \label{18}
\end{equation}
which, for the Hamiltonian (\ref{122}) under consideration, are
\eqb
\mathbf{\dot{r}}\otimes \openone_{2\times 2 } &=& \left(\mathbf{p}+3\,\kappa\,
  {\bf s}  \right) \otimes \openone_{2\times 2 }  + 2 \left(\mathbf{r}
  \times {\bf s} \right) \otimes \sigma_3,
\nonumber
\\
\mathbf{\dot{p}}     \otimes  \openone_{2\times 2 }    &=&     -     \left(
  \mathbf{r}+3\,\theta\, {\bf s} \right) \otimes \openone_{2\times 2 }
+ 2 \left(\mathbf{p} \times {\bf s}\right) \otimes \sigma_3,
\label{19}
\\
\mathbf{\dot{s}} \otimes \openone_{2\times 2 } &=& 3 \left( \kappa \mathbf{p} +
  \theta \mathbf{r} \right) \times \,{\bf s} \otimes \openone_{2\times 2 }
  -3 \left(\mathbf{L}\times {\bf s} \right) \otimes \sigma_3.
\nonumber
\eqf

These equations imply the conservation of the total angular
momentum. Indeed, from the first two equations in (\ref{19}) one gets
\[
\mathbf{\dot{L}}\otimes  \openone_{2\times  2   }  =  -3\left(  \kappa
  \mathbf{p}  +  \theta  \mathbf{x}\right)  \times \,{\bf  s}  \otimes
\openone_{2\times  2  } +  3  \left(\mathbf{L}\times  {\bf s}  \right)
\otimes \sigma_3
\]
and using the third equation in (\ref{19}) one finds
\bb
\frac{d}{dt}  \left( {\bf L} + {\bf s}\right) =0,
\ee
the  aforementioned result.

\medskip

On the other hand, the  different terms  appearing  in the  Heisenberg
equations of  this extension of the
three-dimensional supersymmetric  harmonic oscillator
are relevant in  different contexts. To see this,  note first that the
Hamiltonian in Eq.\ (\ref{122}) is block-diagonal, acting each block on
the corresponding two-component spinor, $\Psi^I$  and  $\Psi^{II}$. But,
as we have seen, only the lower component has a normalizable ground state at zero
energy.
The equations of motion restricted to this sector reduce to
\eqb
\mathbf{\dot{r}}&=& \left(\mathbf{p}+3\,\kappa\,{\bf s} \right)
+2 \left(\mathbf{r}\times {\bf s} \right), \nonumber
\\
\mathbf{\dot{p}}&=& - \left(\mathbf{r}+3\,\theta\,{\bf s} \right)
+2 \left(\mathbf{p}\times {\bf s}\right), \label{20}
\\
\mathbf{\dot{s}}  &=& 3\left(\kappa\mathbf{p}+\theta \mathbf{r}\right)
\times \,{\bf s}
-3 \left(\mathbf{L}\times {\bf s} \right).  \nonumber
\eqf
Here, the last three terms and on the right sides are simply corrections due to supersymmetry
and they are present regardless of the non-commutativity of spin.  The
couplings  ${\bf  r}\times  {\bf  s}$  and  ${\bf  p}\times  {\bf  s}$
correspond to magnetic dipolar and Rashba forces respectively.

The terms coming  from the noncommutative nature of  the problem open
the doors for interesting cases. Consider for example the algebra (\ref{4})
with  $\kappa \neq  0$  and  $ \theta=0$. In  this  case, the  relevant
commutator \cite{grafeno}
\[
[\hat{p}_i,\hat{p}_j] = i \kappa^2 \epsilon_{ijk} \hat{s}_k,
\]
is the analog  of the Landau problem for a  generator $SU (2)$ instead
of $U (1)$. In fact, the  part of the Hamiltonian corresponding to the
free particle, once the shift (\ref{5}) is performed, turn out to be
\[
H= \frac{1}{2m} \left( {\bf p} + \frac{\kappa}{2} \sigmavec \right)^2,
\]
which describes the motion of a particle of mass $m$ interacting with
a chromo-magnetic constant field in the sense that, in the field intensities tensor $F_{\mu\nu}$, with
\[
F_{12} = \partial_1 A_2 - \partial_2 A_1 -i[A_1,A_2],
\]
the choice $A_1 =\kappa \sigma_1$ and $A_2=\kappa\sigma_2$ leads to
\bb \label{gra}
F_{12}= \kappa^2 \sigma_3.
\ee
The Hamiltonian in Eq.\ (\ref{gra}) has been recently employed to model the graphene \cite{grafeno}.

\bigskip

\noindent\textbf{Acknowledgements}: We thank J. C. Retamal by useful discussions. This  work was supported by grants from CONICET (PIP 01787), ANPCyT (PICT 00909) and UNLP (Proy.~11/X492), Argentina, and from FONDECYT-Chile grant-1095106, 1095217, 1100777 and Proyecto Anillos ACT119.


\begin{thebibliography}{99}

\bibitem{nikita} See {\it e.g} M. R. Douglas and N. Nekrasov, {\it Rev. Mod. Phys.} {\bf 73}, 977 (2001)
\bibitem{landau} See L. D. Landau and E. M. Lifshitz, Quantum Mechanics, Vol III, Pergamon, 1962.
\bibitem{wil}   F. Wilczek, {\it Fractional Statistics and Anyon Superconductivity}, World Scientific Anyon Superconductiviy (1990).
\bibitem{np} V. N. Nair and A. P. Polychronakos, {\it Phys. Lett.} {\bf B505}, 267 (2001).
\bibitem{some} The literature  is  very extensive,  some papers  are:  G.   V.  Dunne,  J.  Jackiw  and   C.  Trugenberger,
{\it Phys.  Rev.}   {\bf  D 41}  661  (1990);  J.   Gamboa, M.   Loewe  and
J. C. Rojas, {\it Phys.  Rev. }  {\bf D64}, 067901 (2001); J.  Gamboa, M.
Loewe, J. C.  Rojas and F. M\'endez, {\it Int. J. Mod. Phys.}
5{\bf  A17},  2555  (2002);  J.    Gamboa,  M.   Loewe,  J.  C. Rojas  and
F. M\'endez, {\it Mod. Phys. Lett.} {\bf A16}, 2075 (2001);  H. Falomir, J.  Gamboa, M. Loewe,
F.   M\'endez and J.  C. Rojas,  {\it Phys.  Rev.}  {\bf  D66}, 045018
(2002);   K.  Bolonek and
P.   Kosinski, arXiv:0704.2538;  A. Kijanka  and P. Kosinski, {\it
  Phys. Rev.} {\bf D70}, 127702 (2004); L.  Mezincescu, [hep-th/ 0007046]; C. Acatrinei, JHEP {\bf 0109}, 007 (2001);
M. Gomes and V. G. Kupriyanov, arXiv: 0902.3252 [math-ph]; O. Bertolami and C. Zarro, arXiv:0908.4196; C. Bastos, O. Bertolami, N. Dias, J. Prata, {\it Phys. Rev. }{\bf D78}, 023516 (2008);   O. Bertolami, J.G. Rosa, C.M.L. de Aragao, P. Castorina and  D. Zappala, {\it Phys. Rev.} {\bf D72}, 025010 (2005); M. Gomes, V. G. Kupriyanov and A. J. da Silva, 
{\it Phys. Rev. D } {\bf 81}, 085024 (2010).

\bibitem{gosh} For anyons  see for instance S. Ghosh, 
{\it Phys. Lett. B } {\bf 338}, 235 (1994);  S. Ghosh,   {\it 
Phys.Rev.} {\bf D51}, 5827 (1995).  



\bibitem{snyder} H. Snyder, {\it Phys. Rev.} {\bf 71}, 38 (1947).
\bibitem{yang} C. N. Yang,  {\it Phys. Rev.} {\bf 72}, 874 (1947)
\bibitem{parmi1} In  a different context, the harmonic oscillator in a similar kind of  non-commutative space   has been discussed  in A. Parmeggiani and M. Wakayama, {\it Proceedings of the National 
Academy of Sciences U.S.A.}, {\bf 98}, 26  (2001);  A. Parmeggiani
{\it Communications in Mathematical Physics} {\bf 279} 285(2008).

\bibitem{gz} J. Gamboa and J. Zanelli, {\it Phys. Lett.} {\bf 165B}, 91 (1985).
\bibitem{gozzi} E. Gozzi, {\it Phys. Lett.} {\bf 129B}, 432 (1984); ibid, {\it Phys. Rev.} {\bf D33}, 3665 (1986).
\bibitem{nos}
  H.~Falomir, J.~Gamboa, J.~Lopez-Sarri\'on, F.~M\'endez and P.~A.~G.~Pisani,
  {\it Phys. Lett.}  {\bf 680B}, 384 (2009)
\bibitem{nos2} A. Das, J. Gamboa, F. M\'endez and F. Torres, {\it Phys. Lett.} {\bf A375}, 1756 (2011).
\bibitem{nos1}  A.~Das, H.~Falomir, M.~Nieto, J.~Gamboa and F.~M\'endez, {\it Phys. Rev.} {\bf  D84}, 045002 (2011).
\bibitem{dressel}  G. Dresselhaus, {\it Phys. Rev.} {\bf 100}, 580 (1955).
\bibitem{grafeno} H. Falomir, J. Gamboa, M. Loewe and M. Nieto, \lq \lq Relation of graphene with a non-commutative geometry", arXiv:1109.6666 [math-ph].


\end{thebibliography}
\end{document}